\documentclass[aps,twocolumn,amsmath,amssymb]{revtex4-1}

\usepackage{graphicx}
\usepackage{natbib}

\def\cm{cm$^{-1}$}
\def\sb{SmB$_6$}
\def\sba{Al Flux-SmB$_6$}
\def\sbp{FZ SmB$_6$-Pure}
\def\sbd{FZ SmB$_6$-Defc}

\begin{document}

\title{Breakdown of the Kondo insulating state in \sb\ by introducing Sm vacancies.}

\author{Michael E. Valentine}
\affiliation{Institue for Quantum Matter and Department of Physics and Astronomy, Johns Hopkins University, Baltimore, MD 21218, USA}

\author{Seyed Koohpayeh}
\affiliation{Institue for Quantum Matter and Department of Physics and Astronomy, Johns Hopkins University, Baltimore, MD 21218, USA}

\author{W. Adam Phelan}
\affiliation{Institue for Quantum Matter and Department of Physics and Astronomy, Johns Hopkins University, Baltimore, MD 21218, USA}

\author{Tyrel M. McQueen}
\affiliation{Institue for Quantum Matter and Department of Physics and Astronomy, Johns Hopkins University, Baltimore, MD 21218, USA}


\author{Natalia Drichko}\email{Corresponding author: drichko@jhu.edu}
\affiliation{Institue for Quantum Matter and Department of Physics and Astronomy, Johns Hopkins University, Baltimore, MD 21218, USA}




\author{Priscila F. S. Rosa}
\affiliation{Department of Physics and Astronomy, University of California, Irvine, California 92697, USA}

\author{Zachary Fisk}
\affiliation{Department of Physics and Astronomy, University of California, Irvine, California 92697, USA}

\begin{abstract}

\sb\ is a proposed topological Kondo insulator where the presence of topological nontriviality can be tuned by variations in the Sm valence. Experimentally, Sm valence can be changed by tuning stoichiometry of \sb.  We show that Raman scattering can detect vacancies down to 1\% of Sm sites in \sb\ crystal by probing the intensity of defect-induced scattering of the acoustic phonon branch at 10~meV. In the electronic Raman spectra of \sb\ at temperatures below 130~K, we observe features developing in A$_{1g}$ and E$_g$ symmetries at 100 and 41~meV which we assign to excitations between hybridized bands, and depressed spectral weight below 20~meV associated with the hybridization gap. With the increased number of Sm vacancies up to 1\% we observe an increase of spectral weight below 20~meV showing that the gap is filling in with electronic states. For the sample with the lowest number of vacancies the in-gap exciton excitations with long lifetimes protected by hybridization gap are observed at 16-18~meV in E$_g$ and T$_{2g}$ symmetries. These excitations broaden as a decrease in the lifetime with increasing number of vacancies and are quenched by the presence of in-gap states at concentration of Sm vacancies of about 1\%. Based on this study we suggest that only the most stoichiometric \sb\ samples have a bulk gap necessary for topological Kondo insulators.

\end{abstract}

\date{\today}
\maketitle

\section{Introduction}

Much resent research is aimed on experimental realization of a new state of matter, topological insulator (TI), where topologically protected metallic surface states appear due to a surface crossing of inverted bands in the bulk. While the TI state was initially proposed for band insulators, it was realized that Kondo insulators also can be a source of topologically invariant surface state \cite{Dzero2010,Dzero2012,Alexandrov2013}. \sb\ is one of the materials which are suggested to show properties of topological Kondo insulator (TKI). This material has been studied extensively due to its mixed valence and Kondo insulating properties. 
A presence of the metallic surface states reveal itself as a plateau in the d.c. resistivity temperature dependence below 5~K which has been attributed to surface conduction \cite{Wolgast2013,Phelan2014,Syers2015}. Interpretation of this metallic surface states varies from TI state\cite{Wolgast2013,Syers2015} to polarity-driven surface states\cite{Zhu2013,Phelan2014}.

The basic requirement for TI state to exist is a presence of  inversion of the energy bands forming the  respective insulating gap. In \sb, a gap opens at the Fermi level due to hybridization between $4f$ and $5d$ electronic bands below  70~K (below 150~K according to Ref.~\onlinecite{Jiang2013}). The size of the hybridization gap has been estimated by several techniques. Optical measurements suggests a gap of 16-19 meV with an impurity band at 3-5~meV \cite{Travaglini1984,Ohta1991,Nanba1993,Gorshunov1999}, photoemission at 18 meV \cite{Jiang2013}, point-contact spectroscopy at 21~meV with an in-gap band at 4.5~meV below the conduction band \cite{Flachbart2001,Frankowski1982}, and  DC resistivity estimates the activation energy at 3.5~meV \cite{Wolgast2013}.  Several band structure calculations predict a gap opening due to hybridization of 4$f$ and 5$d$ orbitals, with  band  inversion  necessary for topological nontriviality at the X-point in the Brillouin zone (BZ) \cite{Antonov2002,Lu2013}. A recent neutron scattering study \cite{Fuhrman2015} gives evidence of band inversion  at around $X$ and $R$ points of the Brillouin zone.

It is proposed that a change in the average Sm valence can cause a transistion from a strong topologicial insulator state to a trivial band insulator.\cite{Alexandrov2013}   An introduction of Sm vacancies through increasing Sm deficiency is a means for shifting  average Sm valence. This   can be achieved in a controlled way along the length of a floating zone (FZ) grown crystal \cite{Phelan2015}.   At the transition between the TI and trivial states the gap should close, but for the situations, where the symmetry of the system is broken\cite{Ezawa2013}. The aim of our work was to follow a change of the bulk gap with a change of average Sm valence tuned by the number of Sm vacancies. To do this we use Raman spectroscopy, which can probe bulk electronic structure through measuring intra- and inter-band  excitations\cite{Devereaux2007,Burstein1971}. Low frequency Raman scattering measurements already proved to be useful in studies of the hybridization gap and in-gap states for Al-flux grown \sb\ samples\cite{Nyhus1995,Nyhus1997}. Here we present a wide energy range Raman study  of \sb\ samples with a variation in number of Sm vacancies, which allows us to get rich information on the  phonon and electronic spectrum of these samples.


We demonstrate that phonon Raman scattering  is a sensitive tool to characterize the presence and relative amount of Sm vacancies in \sb\ crystals. In the electronic Raman response at low temperatures for all the samples we observe features associated with the excitations over hybridization gap. We show that the presence of even 1\% of Sm vacancies affects the hybridization gap by filling it with in-gap states. In samples with a negligible number of Sm vacancies, we observe features originating from the in-gap excitonic level.\cite{Alekseev1993,Fuhrman2014,Fuhrman2015} The exciton feature is quenched in the samples with about 1\% Sm vacancies due to the presence of the in-gap impurity states.


\section{Experiment}
\label{sec:exp}

\subsection{Crystal growth}
\label{ssc:crystal}

In this study we used single crystals of \sb\ grown by  Al flux and floating zone (FZ) techniques. Al flux-\sb\ samples where grown by standard Al flux technique. For our study, we selected a crystal, \sba, found from Raman measurements to have the fewest Sm vacancies and comparable linewidths of boron Raman active phonons to FZ samples. A value of the line width of boron phonons as a parameter to characterize the quality of \sb\ samples and the degree of structural variation within Al flux grown samples will be discussed elsewhere~\cite{Valentine2016}.

FZ single crystals of \sb\ were grown using the optical floating zone technique~\cite{Phelan2014, Phelan2015} and are representative of "typical" \sb\ crystals described in these papers. A previous study shows increasing presence of Sm vacancies along the length of a single FZ crystal due to vaporization of the rod materials into a Sm rich mixture and can be characterized by a systematic decrease in lattice parameters. For our study we used two samples, cut from the most stoichiometric (\sbp) and most Sm deficient (\sbd) end of the rod. Based on powder diffraction measurements of the lattice parameters in comparison to previous results for non-stoichiometric \sb, we estimate the most deficient sample (\sbd) to be about 1\% deficient \cite{Phelan2015}. Magnetization measurements for the two FZ-grown samples did not show a significant difference in average magnetic moment which could arise from larger differences in the number of Sm-vacancies \cite{Kasuya1977}.



\subsection{Raman measurements}
\label{sec:raman}

Raman measurements were performed using a Horiba Jobin-Yvon T64000 triple monochromator spectrometer in the pseudo-Brewster angle geometry for energies from 2.5~meV (20~\cm) to 500~meV (4000~\cm) with resolution up to 0.25~meV (2~\cm). A Coherent Ar$^+$ laser was used as a source for excitation light with wavelengths 488~nm and 514~nm which was focused on the samples with a spot size of 50 by 100~$\mu$m.  Penetration depth of the light at 514 nm is estimated to be of the order of 100 nm. The measurements were performed on cleaved surfaces which were exposed to atmosphere before the measurements. No Raman evidence of samarium oxide which typically appears on the surface of the samples exposed to air  was detected in the measured Raman spectra.

Measurements were performed over a temperature range of 10 to 300~K using a Janis ST-500 cold finger cryostat with samples affixed to the cold finger using silver paint. Laser heating was estimated to be 10~K at 10~mW, and the power was reduced to reach the lowest temperatures. All spectra were corrected by the Bose-Einstein thermal factor. To compare the results for different samples the spectra were normalized on the intensity of the 1143 \cm\ phonon to compensate for the small differences in intensity due to the variation in the quality of the cleaved surfaces.

\sb\ has $Pm\bar{3}m$ cubic symmetry of the unit cell, O$_h$ point group symmetry. The crystals were oriented using X-ray diffraction and polarization-dependent Raman scattering measurements. The temperature dependent measurements were performed in $(ab)$ plane with the orientations of the electrical field of the incident light $e_i$ and electrical vector of the scattered light $e_s$ listed in Table~\ref{Geo}. The large acceptance angle of the analyzing optics results in some additional signal from other polarizations. In the table we also present the irreducible representations of the $O_h$ point group probed in these polarizations. Of these, E$_g$ corresponds to B$_{1g}$, and T$_{2g}$ corresponds to B$_{2g}$ irreducible representations of D$_{4h}$ group, for which calculations of electronic Raman scattering were performed\cite{Freericks2001,Devereaux2007}.


\begin{table}
  \begin{tabular}{|c|c|c|c|}
     \hline
Polarization & $e_i$, $e_s$ & Symmetry (O$_h$) \\
             & geometry     &                   \\ \hline
 $(x,x)$    & $c(aa)\bar{c}$                                         &  A$_{1g}$+E$_g$             \\
 $(x,y)$    & $c(ab)\bar{c}$                                         &  T$_{2g}$                   \\
 $(x',x')$  & $c(\frac{a+b}{\sqrt(2)},\frac{a+b}{\sqrt(2)})\bar{c}$  &  A$_{1g}$+1/4E$_g$+T$_{2g}$ \\
 $(x',y')$  & $c(\frac{a+b}{\sqrt(2)},\frac{a-b}{\sqrt(2)})\bar{c}$  &  3/4E$_g$                   \\
     \hline
   \end{tabular}
  \caption{Polarizations of the measured Raman scattering spectra of \sb, the geometry of the measurements,  and the probed irreducible representations for each polarization.}\label{Geo}
\end{table}


\section{Results}

The Raman spectra of \sb\ consist of relatively narrow phonons peaks superimposed on the electronic background. Most of the features of the phonon spectrum have been previously studied \cite{Morke1981,Lemmens1995,Nyhus1995,Nyhus1997,Ogita2003,Ogita2005}. In Sec.~\ref{sec:phonons}, we discuss how previously unidentified defect-induced phonon scattering can be used to extract the information on Sm vacancies in the studied samples of \sb. The electronic Raman response for the samples with different numbers of Sm vacancies is discussed in Sec.~\ref{sec:hgap}.

\subsection{Phonons}
\label{sec:phonons}

\begin{figure}
	\includegraphics[width=8cm]{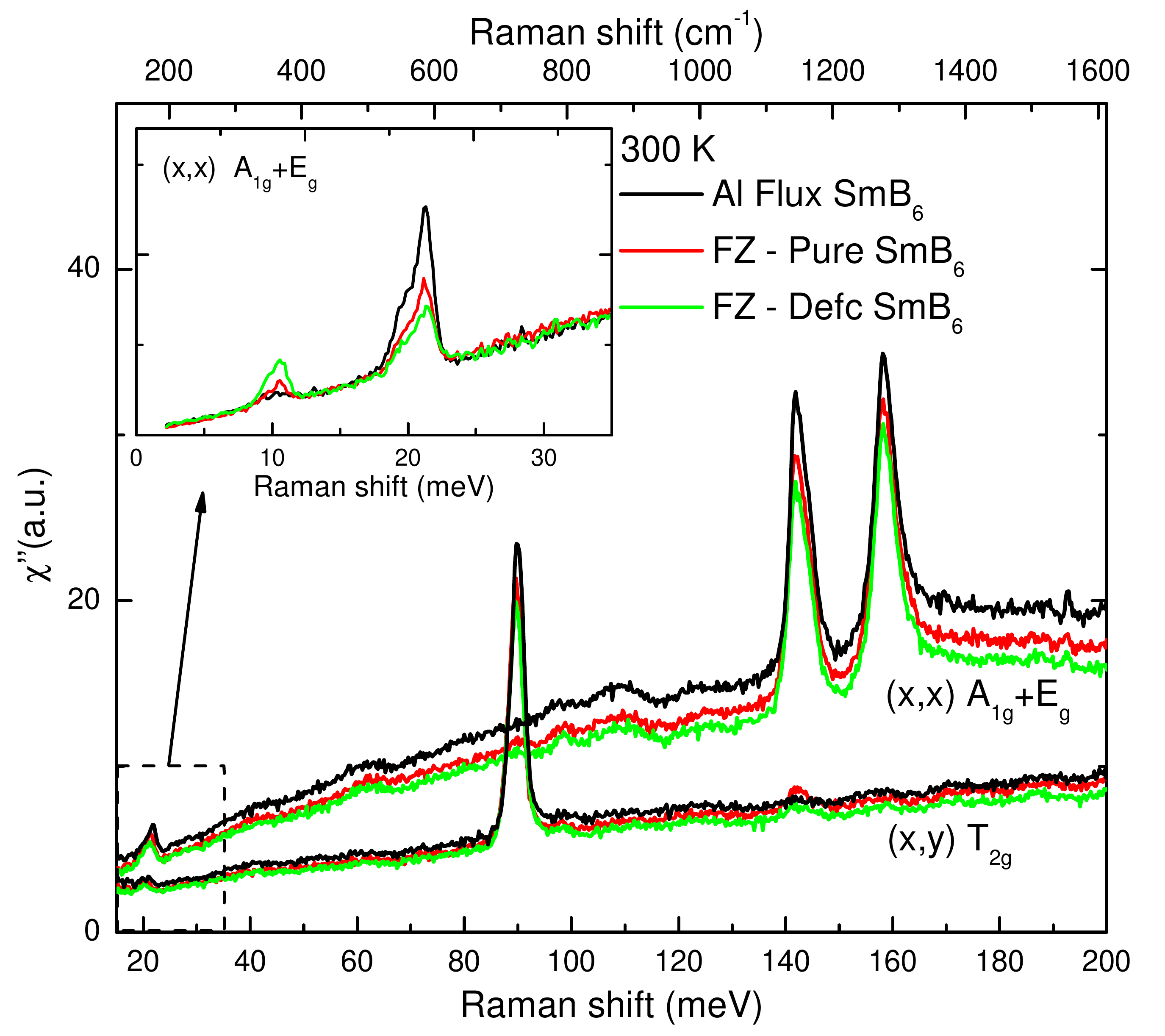}
	\caption{Room temperature Raman spectra of the three studied SmB$_6$ samples with increasing number of Sm vacancies (\sba, \sbp, \sbd) in $(x,x)$ and $(x,y)$ polarizations. The 3 first-order Raman active phonons appear at 89.6~meV ($T_{2g}$), 141.7~meV ($E_g$), and 158.3~meV ($A_{1g}$) are superimposed on a broad continuum of electronic scattering. Inset shows low-frequency $(x,x)$ spectra of the samples. Two symmetry forbidden peaks appear at 10~meV and 21~meV correspond to defect-induced and two-phonon scattering, respectively.}
	\label{fig:RT}
\end{figure}

In the cubic unit cell of \sb, the Sm ions are located at a center of inversion symmetry, and thus phonons involving this site are forbidden in the first order Raman spectrum. The three symmetry allowed Raman phonons seen as the intense relatively narrow features (see Fig.~\ref{fig:RT}) are the $T_{2g}$ symmetry phonon at 89.6~meV (723~\cm), $E_g$ at 141.7~meV (1143~\cm), and $A_{1g}$ at 158.3~meV (1277~\cm) exclusively involve motion of the atoms within the B$_6$ octahedra \cite{Ogita2003}. They are observed in polarizations corresponding to their symmetries and are well-known from other vibrational Raman studies of \sb\ crystals\cite{Morke1981,Nyhus1995,Nyhus1997,Ogita2003,Ogita2005}. The widths of the phonons are 2-4~meV which emphasizes the role of valence fluctuations and disorder in the crystals~\cite{Shuker1970}. While Al-flux grown samples show variations among the first order B$_6$ octahedra phonon energies of 1~meV and linewidths 0.8~meV \cite{Valentine2016} for our measurements we chose the sample (\sba) with the lowest line width of boron phonons, which was an evidence of a minimized disorder in the sample. For this sample, the positions of the boron phonons coincide with
that of the FZ-grown samples.

The phonon features at 10 and 21~meV (see Fig.~\ref{fig:RT}) observed strongest in $(x,x)$ polarization are assigned to phonons associated with Sm motion, which are Raman silent for the $Pm\bar{3}m$  symmetry of the unit cell. Previously both features were associated  with two-phonon scattering.

The 21~meV feature has two components, the sharper peak at 21.9~meV that has maximum intensity in $(x,x)$ polarization, and a wider polarization-independent component with a maximum at 20.3~meV. The intensity of  this feature as a whole decreases on going from Al flux SmB$_6$ to FZ-Defc-SmB$_6$ sample (see inset in Fig.~\ref{fig:RT}), but for all of the samples the feature decreases in intensity on cooling. The feature was previously attributed to the two-phonon scattering from acoustic phonons\cite{Lemmens1995,Nyhus1997}, based on its temperature dependence. Optical phonon branches, as well as a non-dispersive  mode of a phonon coupled to the  valence fluctuations are observed by neutron scattering in the same energy range \cite{Alekseev1989,Alekseev2015}, however,  more experimental data is necessary to understand if those affect the Raman intensity.


The assignment of the feature at 10~meV   to two-phonon scattering \cite{Ogita2003, Ogita2005}  does not account for the temperature dependence which follows a first-order thermal factor $\frac{1}{1-e^{-\frac{\hbar \omega}{kT}}}$, and the lack of singularities in phonon density of states at half this energy necessary for strong two-phonon scattering \cite{Alekseev1989}. We instead attribute this feature to acoustic phonons, which become Raman active due to local symmetry breaking induced by the presence of Sm defects. This loss of translational invariance allows light scattering from all points within the BZ~\cite{Shuker1970}. The relevant acoustic phonons with  energies around 10~meV and a flat dispersion over the latter half of the BZ  are observed in neutron scattering experiments\cite{Alekseev1989}. The flat dispersion of this phonon is responsible for the relatively small line width of the Raman feature.

The Sm defect-induced phonon at 10~meV shows an increase in spectral weight with increasing Sm deficiency between the two FZ-grown samples (Fig.~\ref{fig:RT}).  While the increase of the number of Sm vacancies between these two samples is estimated to be less than 1\%, the difference in the intensity of the Sm defect-induced phonon at 10~meV is about 1.5 times and can be easily detected. This shows that Raman scattering can be effectively used to characterize the number of Sm vacancies in \sb\ samples. The intensity of 10~meV phonon has nearly zero spectral weight for the best \sba\ sample, demonstrating that the sample has the lowest number of Sm vacancies. The intensity of this Sm symmetry-forbidden phonon allows us to order the three studied samples by increasing number of Sm vacancies from \sba to \sbp, and further to \sbd. As the next step we follow the temperature dependence of electronic Raman scattering within this range of samples.


\subsection{Electronic Raman scattering}
\label{sec:hgap}

\begin{figure} [h!]
	\includegraphics[width=9cm]{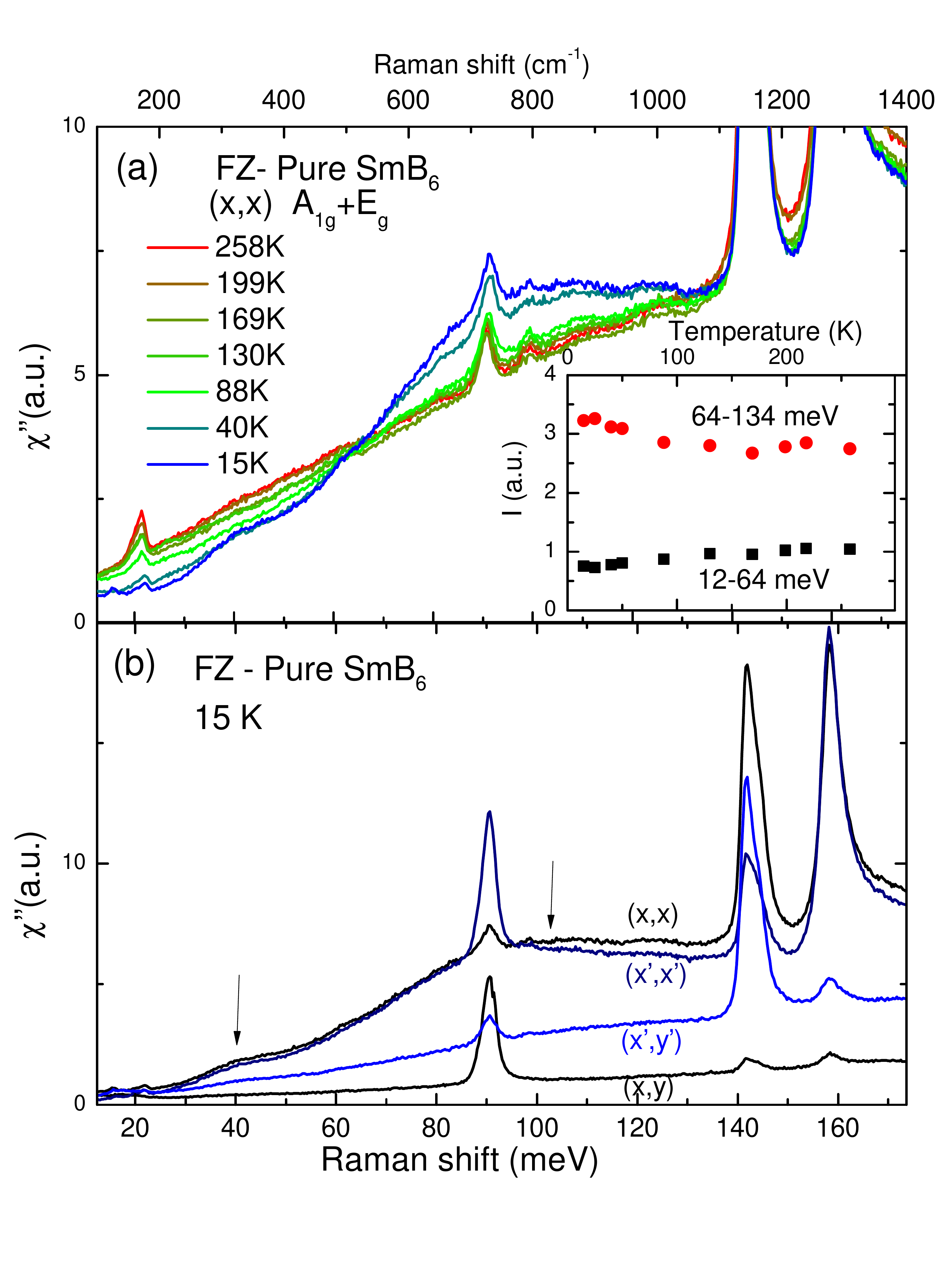}
	\caption{(a)Temperature dependence of Raman spectra of the \sbp\ sample cooled from 300~K to 15~K in $(x,x)$ polarization. Note redistribution of the spectral weight which occurs below 130~K to the frequencies above 100~meV, and below 50~K to the frequency range above 34~meV. The inset shows a temperature dependence of spectral weight $I(T)=\int^{\omega_0}_{\omega_1}\chi''(T,\omega) d\omega$ below ($\omega_0$ = 12 meV,    $\omega_1$ = 64 meV) and above ($\omega_0$ = 64 meV and  $\omega_1$ = 134 meV) the isobestic point. (b)   Raman spectra of  \sbp\ sample at 15~K in $(x',y'), (x',x'), (x,y)$ and $(x,x)$ polarizations, see Table~\ref{Geo}. The temperature dependent response is most intense in $(x',x')$ and $(x,x)$ polarizations, suggesting that it belongs to A$_{1g}$ symmetry. }
	\label{fig:TDep}
\end{figure}

In the Raman spectra of all three samples at 300~K, we observe electronic backgrounds which linearly increase in intensity with energy up to about 150~meV (1200~\cm) and stays constant at higher energies (see Fig.~\ref{fig:RT}). The background is observed in all four measured polarizations, though it is considerably weaker in $(x,y)$. This background is present in the spectra excited with 488 nm line as well, which suggest that it originates from electronic Raman scattering \cite{Burstein1971}.

The changes observed in the Raman spectra of all the samples on cooling from 300 to 15~K are illustrated  by the temperature dependence the response of the \sbp\ sample in $(x,x)$ polarization presented in Fig.~\ref{fig:TDep} (a). On decreasing temperatures below 130~K, we detect a spectral weight shift to frequencies above an isosbestic point of 64~meV. The resulting feature with a maximum at about 100~meV continues to develop down to 20~K. We can follow the temperature dependence of the high-frequency feature by following the temperature dependence of the  spectral weight $I(T)=\int^{\omega_0}_{\omega_1}\chi''(T,\omega) d\omega$ below ($\omega_0$ = 12 meV,    $\omega_1$ = 64 meV) and above ($\omega_0$ = 64 meV and  $\omega_1$ = 134 meV) the isobestic point,  $\chi''(T,\omega)$ is Raman intensity in arbitrary units. Another redistribution of the spectral weight occurs at temperatures below 50~K, resulting in a band at 41~meV with further suppression of the spectral weight below 34~meV. This lower-frequency effects are in general agreement with Ref.~\cite{Nyhus1995,Nyhus1997}, while the feature at about 100~meV was not yet discussed.  The total spectral weight of the spectra below 134~meV (the sum of the two parts) is conserved, as expected for a system where a metal-insulator transition is driven by electronic correlations~\cite{Freericks2005}.

As seen from polarization dependence of the spectra at 15~K (Fig.~\ref{fig:TDep} (b)), both features have the highest intensity in $(x,x)$ and $(x',x')$, with somewhat lower intensity at the same frequencies observed in $(x',y')$. This shows that both features appear in A$_{1g}$ and E$_g$ symmetries at the same energies. In $(x,y)$ polarization (T$_{2g}$ symmetry, see Sec.~\ref{sec:raman}) the Raman response in this frequency range is low and basically temperature independent, and neither of these two features are observed at 15~K.

\begin{figure}
	\includegraphics[width=9cm]{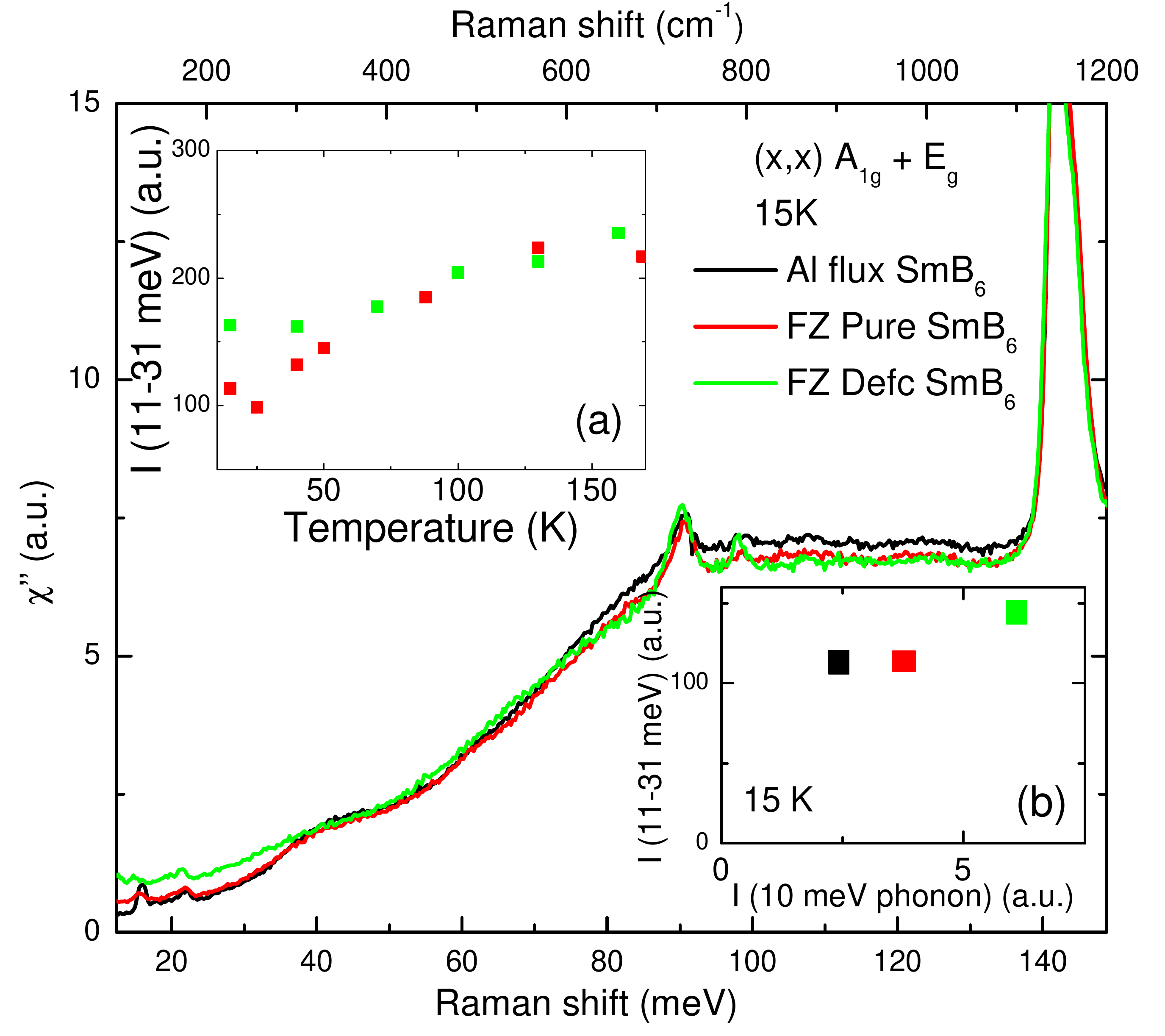}
	\caption{Low-temperature Raman spectra of \sba, \sbp, and \sbd\ samples at 15~K in $(x,x)$ polarization. Note an increase of in-gap intensity and smearing of 41~meV feature with the increase in Sm vacancies.   (a) Temperature dependence of the spectral weight $I(T)=\int^{\omega_0}_{\omega_1}\chi''(T,\omega) d\omega$  below 31.5 meV in \sbp\ sample (red dots) vs \sbd\ sample (green dots). The difference becomes apparent below 50 K, where the 41~meV feature starts to develop in the spectra.   (c) Spectral weight below 31.5 meV plotted against the intensity of the defect phonon. Note the increase of the low frequency spectral weight with the increase of the number of Sm vacancies.}
	\label{Allat15K}
\end{figure}

A similar changes of the spectra on cooling  are observed in the other samples. We compare the $(x,x)$ spectra at 15~K for the samples with different concentrations of Sm vacancies in Fig.~\ref{Allat15K}. The position and intensity of the feature at 100~meV is the same for all measured samples. The feature at 41~meV gets smeared with an increase of the number of vacancies leaving some spectral weight at low frequencies. We follow this as the decrease of the spectral weight on cooling $I(T)=\int^{\omega_0}_{\omega_1}\chi''(T,\omega) d\omega$ between $\omega_0$ = 11 meV, $\omega_1$ =31.5 meV for \sbp\ (red squares) and \sbd\ (green squares) in the inset (a) in Fig.~\ref{Allat15K}.  The spectral weight shows identical dependence on temperature in both samples down to approximately 50~K. Below this temperature no major changes occur in the low frequencies range for the \sbd\ sample, while  the further decrease of the low frequency spectral weight is observed in \sbp\ sample. The resulting correlation between the number of vacancies estimated as the intensity of the 10~meV phonon $I$(10~meV~phonon) and the low frequency spectral weight $I$(11-31~meV) is shown in the inset (b) of Fig.~\ref{Allat15K}. With the decrease of the number of vacancies the low frequency spectral weight decreases.


\begin{figure}
	\includegraphics[width=9cm]{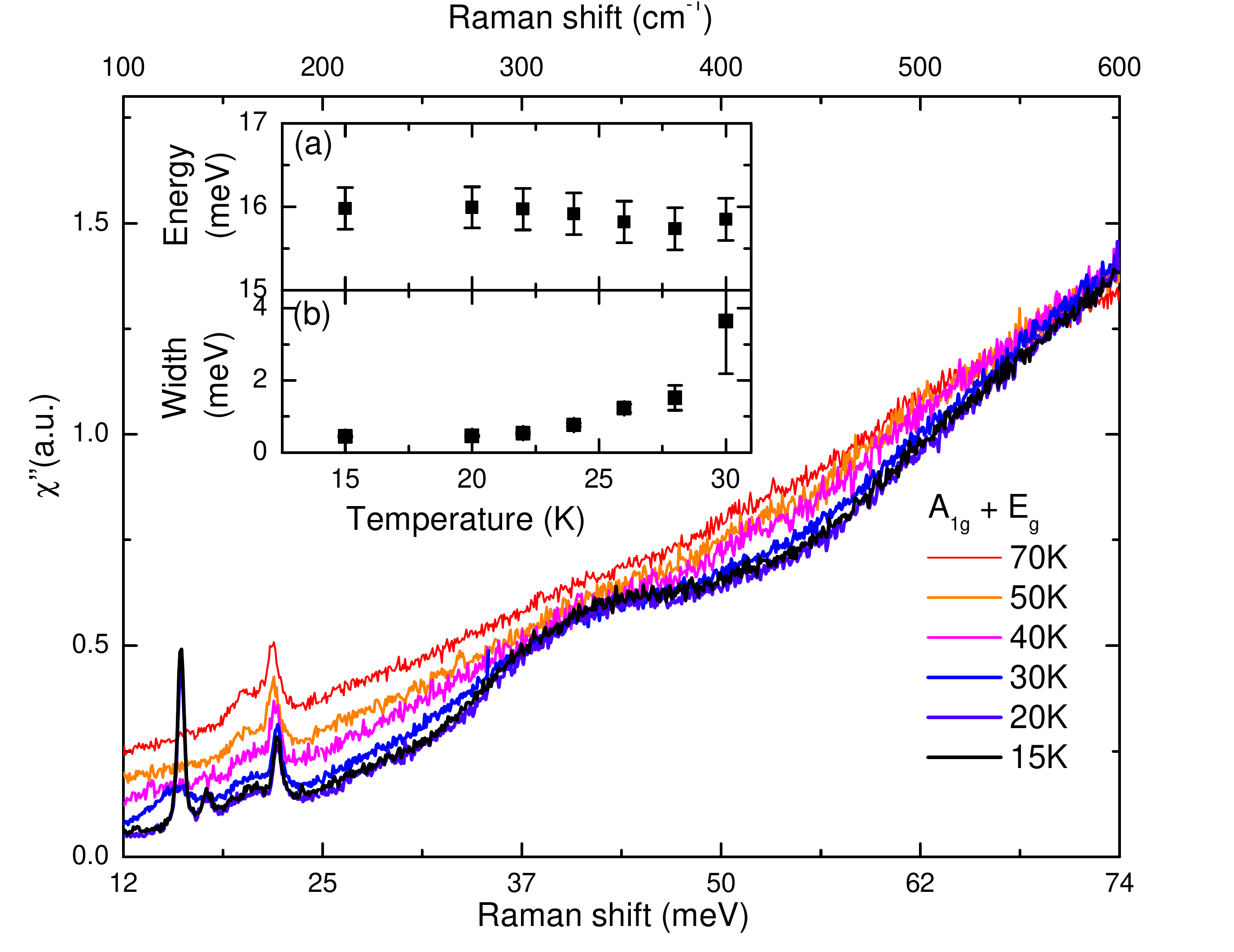}
	\caption{Temperature dependence of the low frequency Raman response of \sba\ sample in $(x,x)$ polarization. The exciton feature appears below 30~K at 16~meV. The inset shows a change of the position and width of the exciton on cooling.}
	\label{fig:exciton}
\end{figure}

\begin{figure}
\begin{center}
	\includegraphics[width=8cm]{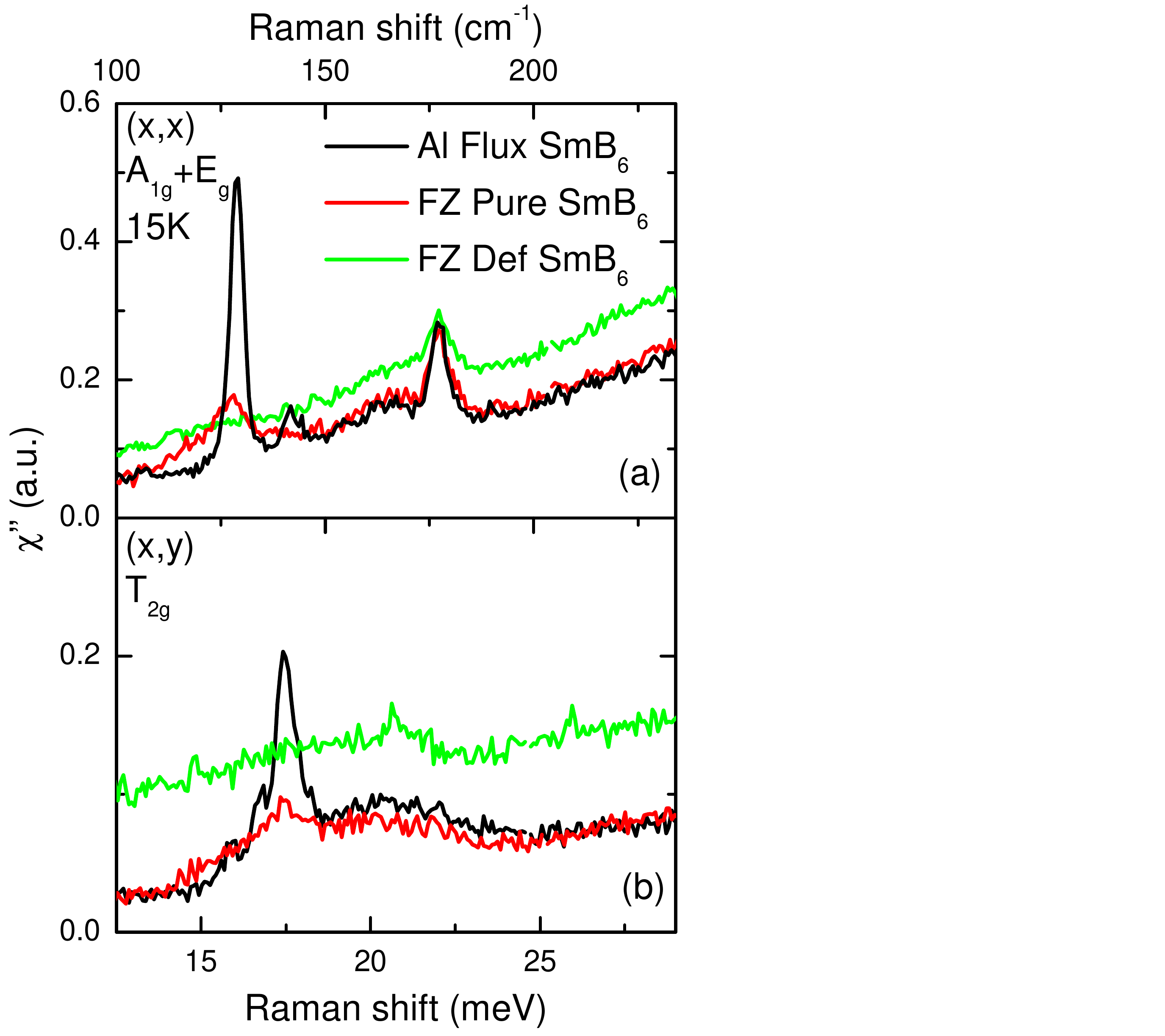}
	\caption{Raman spectra in the frequency range of the exiton feature at 15~K for the measured samples, \sba, \sbp, and \sbd\ in $(x,x)$ polarization (a) and $(x,y)$ polarization (b).}
	\label{fig:AllExcitons}
\end{center}
\end{figure}

While the low frequency spectral weight has similar values in \sbp\  and \sba\,   in the spectra of the \sba\ sample  at temperatures below 30~K the intensity is concentrated in the   narrow features of in-gap excitations (see  Fig.~\ref{fig:exciton} and Fig.~\ref{fig:AllExcitons}) observed both in $(x,x)$ and $(x,y)$ polarizations. Table~\ref{tbl:excit} lists all of the in-gap features, including those in $(x,y)$ polarization, and their symmetry assignment following  Ref.~\cite{Nyhus1997}. The temperature behaviour followed in $(x,x)$ polarization shows that the sharp peak forms when the spectral weight at low frequencies becomes sufficiently low, the frequency of the feature stays the same 16.0~meV, while the width decreases with temperature till it reaches the value of about 0.5 meV (see Fig.~\ref{fig:exciton} (b)).

In the \sbp\ sample lower intensity and wider (1.7~meV) in-gap feature at 15.6~meV is superimposed on even wider (4.7~meV) background. The in-gap excitations are not observed in the spectra of the \sbd, where higher electronic scattering intensity is present at frequencies within the hybridization gap (Fig.~\ref{fig:AllExcitons}).

\begin{table}
\begin{tabular}{|c|c|c|c|c|c|}
  \hline
 Polariz.         &  Symm.   &\multicolumn{2}{|c|}{Al Flux} & \multicolumn{2}{|c|}{FZ Pure} \\
                  &          & Freq       & Width           & Freq       & Width            \\ \hline
  $(x,y)$         & $T_{2g}$ & 15.9 meV   & 0.7 meV         &            &                  \\
  $(x,x)$         & $E_{g}$  & 16.0 meV   & 0.5 meV         & 15.6 meV   & 1.7 meV          \\
  $(x,y)$         & $T_{2g}$ & 16.7 meV   & 0.7 meV         &            &                  \\
  $(x,y)$         & $T_{2g}$ & 17.5 meV   & 0.7 meV         & 17.4 meV   & 1.5 meV          \\ \hline
\end{tabular}
\caption{Polarization dependence, symmetry, frequency and width of the exitonic features observed in the samples with the smaller numbers of Sm vacancies (\sba\ and \sbp) at 15~K.}\label{tbl:excit}
\end{table}

\section{Discussion}

The excitations revealed in the low temperature Raman spectra are of electronic origin, and appear due to hybridization of $5d$ and $4f$ orbitals, as suggested by their temperature dependence and energy scale. The interpretation of the electronic Raman spectra can be made from two points of view. The low temperature electronic Raman spectra can be compared to the band structure calculations, which take into account electronic correlations and hybridization of $5d$ and $4f$ bands, and reproduce the hybridization gap and band inversion  in  \sb \cite{Antonov2002,Lu2013,Alexandrov2013}. On the other hand, the change of the low-frequency electronic Raman spectra of \sb\ on cooling, the formation of hybridization gap,  and polarization dependence of the spectra can be compared to the calculations of electronic Raman spectra for a system with localization effects originating from electronic correlations\cite{Freericks2005}.

The shift of the spectral weight to the higher frequencies on cooling indicates the formation of the insulating gap due to correlation effects, as was already discussed in connection with \sb\ low frequency Raman spectra \cite{Freericks2005}.  In our measurements, for all the samples the spectral weight is  conserved above  105~meV (850~\cm), this cutoff  indicating the energy range  associated with the formation of the hybridization gap.

Of the band structure calculations, in particular that performed by LDA+Gutzwiller method presented in Ref.~\onlinecite{Lu2013} provides bands energies which correspond well to the   electronic excitations observed in our data at low temperatures. The calculations show the hybridized $5d$ and $4f$ bands, where $4f$ $j = 5/2$ band splits into two $\Gamma^f_7$ and one $\Gamma^f_6$ bands resulting in band inversion in the vicinity of the X point of the B.Z.  and a formation of a semiconductor-like gap. The possible interband excitations would have energies close the 41 and 100~meV observed in our experiment (Fig.~3 in Ref.~\onlinecite{Lu2013}). Interestingly,   optical conductivity also shows a broad peak at around 100 meV \cite{Nanba1993}. In inversion symmetric crystals Raman measurements should only probe electronic transitions from bands of the same parity while infrared is limited to bands of different parity. Here the mixed parity of the hybridized {\it 5d-4f} orbitals in \sb\ apparently allows this transition to appear in both measurements.

Depending on the measured polarization, Raman scattering probes electronic excitations at different parts of the B.Z. \cite{Devereaux2007}.  The  response in A$_{1g}$ probes the excitations over the whole B.Z. The 100 meV and 41~meV  features appear with the highest intensity in  A$_{1g}$ symmetry.  According to the calculations\cite{Lu2013}, the dispersions of the relevant bands along the $\Gamma$-X direction are relatively flat, and thus would result in peaks in the Raman response.   E$_g$ symmetry  which corresponds to B$_{1g}$ for D$_{4h}$ point group probes the transitions around the $X$ point and is expected to show Raman response due to correlation effects \cite{Freericks2001}.  We do observe both features in E$_g$ response at similar frequencies as in A$_{1g}$ within the precision of the measurements, which is in agreement with the flat band dispersion.

Following the interpretation of the low temperature spectra within the band structure calculations \cite{Lu2013}, we would expect to observe a predicted much larger gap in T$_{2g}$ symmetry, which probes the $M$ point of the BZ. However, in this symmetry the Raman electronic background has   low intensity  which does not change  with  temperature  (see Fig. 2 (b)). On the other hand, the Raman response due to correlation effects in T$_{2g}$ which can be projected on B$_{2g}$ for D$_{4h}$ symmetry is predicted to be zero\cite{Freericks2001} in approximations of cosine bands.  This absence of temperature-dependent response in T$_{2g}$ can be taken as an evidence of the defining importance of the electronic correlations for the effects observed in \sb\ spectra and a limitation of the band-structure approach.

The two features in the Raman spectrum at 100 and 41~meV associated with hybridization of the $5d$ and $4f$ orbitals start to develop at the relevant temperatures, 130 and 50~K. At low temperatures, the 100~meV feature has equal intensity in all three samples, as seen from Fig.~\ref{Allat15K}. The differences in the low-frequency spectral weight  between \sbp\ and \sbd\ samples appear only below 50~K (Fig.~\ref{Allat15K}, inset (a)).
The depressed  Raman intensity below roughly 30~meV in A$_{1g}$ and E$_g$ symmetries in \sba\ and \sbp\ samples is a result of the opening of the hybridization gap. By evaluating in-gap spectral weight in the gap (Fig.~\ref{Allat15K}(b)) we cannot distinguish the two samples with the least number of vacancies,  \sba\ and \sbp\, while we observe a somewhat higher intensity in the spectra of \sbp\ sample at low frequencies.   With the increase of the number of Sm vacancies for the \sbd\  sample  an increase of the in-gap spectral weight and the respective decrease of the 41~meV band are detected. This increase of the amount of the in-gap spectral weight suggests that the hybridization gap is not opened completely in \sbd\ sample.

At 15~K we observe the sharp in-gap excitations at about 16~meV in \sba\, while in \sbp\ spectra the same spectral weight is spread over  wide frequency range in the gap, with a wide peak in the same range (Fig.~\ref{fig:AllExcitons} ).
The excitation at 16~meV can be interpreted in terms of in-gap exciton.  The respective excitonic level is proposed to be formed   by electrons of hybridized bands in the gap as a result of strong electron-electron correlations and is protected from decay by the hybridization gap \cite{Fuhrman2014}. The excitonic feature was observed in neutron scattering measurements at finite momentum transfers with scattering intensity at the $X$ and $R$ high symmetry points of the BZ\cite{Fuhrman2015}. The most intense feature observed in Raman spectra belongs to E$_g$ symmetry, which probes the excitation at the $X$ point of the BZ\cite{Fuhrman2015}. Raman probes zero momentum transfer direct transitions, and thus the energies of these excitation of 16~meV is higher than that observed in neutron scattering at 14~meV at $X$ point.

The multiple features of T$_{2g}$ symmetry (Fig.~\ref{fig:AllExcitons}, Table~\ref{tbl:excit}) cannot be assigned to pure electronic response.  The previous work\cite{Nyhus1995,Nyhus1997} proposes an alternative model for the   multiple excitonic features which appear as a result of a splitting of crystal field levels by coupling with phonons\cite{Thalmeier1982}.  According to the model,  the feature at 16 meV in E$_g$ symmetry has dominant  electronic contribution, while multiple exciton-related features in T$_{2g}$ symmetry have dominant phonon contribution \cite{Nyhus1997}.

The low line width of the excitonic features observed in the \sba\ sample (0.5 meV) is evidence of an exceptionally long life time protected  by the hybridization gap. The widening of the feature on the increase of the number of vacancies in \sbp\ shows that some electronic state are present within the   gap energies and  lead to the exciton decay. Respectively, no exciton features are detected in the spectra of \sbd\, since the gap is not fully opened in this material.

Our Raman results for FZ-grown samples are mirrored by transport measurements which show a decrease of the metallic-like plateau in resistivity at low temperatures and more insulating behavior for the \sbp\ samples compared to \sbd \cite{Phelan2015}. Such results emphasize the dramatic effect the presence of Sm vacancies has on the bulk hybridization gap   necessary for the existence of KTI. While in the most stoichiometric samples (\sba\ in this study) we find the hybridization gap fully opened and detect the presence of the in-gap exciton, which can be an evidence of the KTI state, even 1 \%   increase  in the number of Sm vacancies due to the growth condition suppresses the development of hybridization gap and can  eliminate the KTI state. The presence of  Sm vacancies can introduce doping, but the effects of disorder also can be important. To probe the exact mechanism of an   effect of  Sm vacancies   on the metallic low-temperature response of \sb\   one needs to use frequency-dependent optical techniques which in contrast to Raman scattering have an ability to directly probe the bulk charge carriers.

\section{Conclusions}

We present Raman scattering study of \sb\ samples with different numbers of Sm vacancies. We show that Raman scattering is an extremely sensitive method to characterise the number of Sm vacancies by the  estimation of the intensity of the Raman-forbidden phonon of Sm at 10~meV, which appears in the spectra due to the local symmetry breaking by Sm vacancies.

In the Raman spectra below 130~K for all the samples we observe a development of the  electronic features at 100 and 41~meV in A$_{1g}$  and E$_g$ symmetries. Based on the recent band structure calculations we assign the features to the excitations between the bands in the electronic structure which appear due to hybridization between $5d$ and $4f$ orbitals.  In turn, our Raman study provides experimental data on electronic structure of \sb\ to support the calculations. In this interpretation, the band at about 41 meV is the excitation across the hybridization gap. While the feature at 100~meV develops equally in all the samples, with an increasing of the number of Sm vacancies up to approximately 1\% the hybridization gap stays filled with states  without a detectable shift in the size of the gap.

For the samples with the low number of Sm vacancies in E$_g$ symmetry, which probes the $X$ point of BZ  we observe a feature of  excitonic excitation at 16~meV.   The extremely low width of 0.5~meV of the exciton feature in \sba\ spectra suggests the extremely long life time of the level. The presence of Sm vacancies lead to a decrease of the exciton life time with eventual decay of   the   exciton  through the electronic states present in the   hybridization gap at   of 1\% of vacancies.

 Our results suggest that an introduction  of 1\% of Sm vacancies while potentially changing Sm average valence has much larger effect on hybridization gap, preventing it from fully opening. The suppression of the hybridization gap by the presence of Sm vacancies can in turn suppress KTI state.

\section{Acknowledgements}

We are grateful to C. Broholm, P. Nikolc, W. Fuhrman,  J. Paglione, and N. P. Armitage for useful discussions. The work at IQM was supported by the
U.S. Department of Energy, Office of Basic Energy Sciences, Division of Material Sciences and Engineering
under Grant No. DE-FG02-08ER46544.

\bibliography{./SmB6}

\end{document}